\documentclass[pra,aps,twocolumn,10pt,showpacs,groupedaddress,superscriptaddress,floatfix,notitlepage]{revtex4-2}
\usepackage{amsmath,amsfonts,amssymb,graphics,graphicx,epsfig,color,times, braket}%,bbm}
\usepackage[utf8x]{inputenc}
\usepackage{color}
\usepackage{bbm, dsfont}
\usepackage{subfigure}
\usepackage{hyperref} 
\usepackage{mathrsfs}
\usepackage{verbatim} 
\begin{document}

\title{Fast charging of quantum batteries assisted by nonlinear couplings}

\author{Y.-J. Shen}
\affiliation{Interdisciplinary Program of Engineering, National Tsing Hua University, Hsinchu 30013, Taiwan}
\affiliation{Institute of Atomic and Molecular Sciences, Academia Sinica, Taipei 10617, Taiwan}

\author{Y.-C. Shen}
\affiliation{School of Electrical and Computer Engineering, Georgia Institute of Technology, Atlanta 30332, Georgia, USA}

\author{C.-H. Chu}
\affiliation{Institute of Atomic and Molecular Sciences, Academia Sinica, Taipei 10617, Taiwan}

\author{H. H. Jen}
\email{sappyjen@gmail.com}
\affiliation{Institute of Atomic and Molecular Sciences, Academia Sinica, Taipei 10617, Taiwan}
\affiliation{Department of Physics, National Central University, Taoyuan City 320317, Taiwan} 
\affiliation{Molecular Science and Technology Program, Taiwan International Graduate Program, Academia Sinica, Taiwan} 
\affiliation{Physics Division, National Center for Theoretical Sciences, Taipei 10617, Taiwan}

\date{\today}
\renewcommand{\r}{\mathbf{r}}
\newcommand{\f}{\mathbf{f}}
\renewcommand{\k}{\mathbf{k}}
\def\p{\mathbf{p}}
\def\q{\mathbf{q}}
\def\bea{\begin{eqnarray}}
\def\eea{\end{eqnarray}}
\def\ba{\begin{array}}
\def\ea{\end{array}}
\def\bdm{\begin{displaymath}}
\def\edm{\end{displaymath}}
\def\red{\color{red}}
\pacs{}
\begin{abstract}
Charger-mediated quantum batteries offer promising opportunities for efficient energy transfer and storage at the quantum level, with potential advantages beyond their classical counterparts. Here, we theoretically demonstrate that nonlinear charger–battery interactions can substantially accelerate the charging of quantum batteries. In the weak-driving regime, sufficiently strong nonlinear coupling can dominate over the effect of dephasing and significantly reduce the charging time. In the strong- and optimal-driving regimes, where the optimal driving strength maximizes the steady-state ergotropy in the linear-coupling regime, nonlinear coupling gives rise to a new optimum in the charging time beyond that of the linear-coupling regime, accompanied by nearly critically damped dynamics. Remarkably, even in the absence of external driving, nonlinear coupling alone can substantially reduce the charging time, providing an alternative route toward fast battery charging. Our results establish nonlinear charger–battery coupling as an effective resource for accelerating quantum-battery charging and broaden the scope of interaction-enabled quantum advantages in energy transfer and storage.
\end{abstract}
\maketitle
%%%%%%%%%%%%%%
\section{Introduction}

Quantum batteries \cite{Alicki2013, Binder2015, Campaioli2017, Campaioli2024, Ferraro2018, Ferraro2026} harness quantum-mechanical principles to envision future mesoscopic devices capable of efficiently transferring and storing energy at the nanoscale. This emerging research direction builds upon foundational studies in quantum thermodynamics \cite{Deffner2019}, which seek to uncover quantum advantages in thermodynamic processes, including faster work extraction \cite{Alicki2013, Hovhannisyan2013} and enhanced heat-engine efficiency \cite{Quan2007, Jaramillo2016, Chand2021, Feyisa2024_1, Feyisa2024_2}. More recently, diverse strategies have been proposed to optimize the performance of quantum batteries, including nonreciprocal charging \cite{Ahmadi2024}, multimode charging \cite{Yang2024}, charging through reservoir engineering \cite{Xu2021, Wang2026} or continuous environmental monitoring \cite{Cenedese2026}, collective charging \cite{Shukla2026, Campaioli2017} with a charger line \cite{Cong2026}, and dephasing-enabled charging \cite{Shastri2025}. These developments highlight the crucial role of dissipation and open-system dynamics \cite{Bruer2002} in quantum batteries, where environmental effects, rather than merely limiting performance allowed in closed systems, can be harnessed as useful resources for enhancing energy storage and work extraction.

Quantum batteries modeled as open quantum systems provide a realistic framework, since dissipation, decoherence, and dephasing are ubiquitous and often unavoidable in physical implementations \cite{Campaioli2024}. Significant progress toward prototype quantum batteries has been achieved across various platforms, including superconducting qubits \cite{Hu2022, Hu2026}, semiconductor quantum dots \cite{Wenniger2023}, organic microcavities \cite{Quach2022}, and nuclear spins \cite{Joshi2022}. Beyond these theoretical and experimental advances, external driving offers an additional means of improving charging performance. In particular, driven quantum batteries \cite{Mondal2022, Rodriguez2024} can exhibit accelerated charging in a kicked-Ising setting \cite{Romero2026}. Despite these developments, nonlinear couplings in charger-mediated quantum batteries remain comparatively unexplored as a resource for enhancing charging performance. Recent studies have demonstrated, for example, that dark states of quantum emitters can be generated through nonlinear interactions in waveguide settings \cite{Karnieli2025, Nadel2026}, which can further facilitate higher-order quantum correlations among the emitters \cite{Handayana2026}. These findings highlight the potential of nonlinear couplings for controlling and manipulating nonclassical correlations, thereby opening a largely unexplored avenue for improving quantum-battery performance and revealing new regimes of quantum advantage.

In this work, we consider a charger-mediated quantum battery and introduce nonlinear couplings between the charger and the battery \cite{Campaioli2024}. Within this open-system setting, we investigate energy transfer and work extraction in the presence of finite dephasing \cite{Shastri2025}. While an optimal dephasing rate can accelerate battery charging in the linearly coupled regime, we find that nonlinear couplings open a distinct route toward even faster charging. In particular, an optimal interplay between nonlinear coupling and dephasing emerges before energy transfer is eventually suppressed at stronger dephasing. We further investigate and compare both driven and undriven scenarios, demonstrating that nonlinear energy exchange between the charger and the battery plays an essential role and can serve as a resource for enhancing battery performance. Our results demonstrate nonlinear coupling as a useful resource for accelerating the charging of quantum batteries and reveal distinct regimes of optimized performance arising from the interplay among coherent driving, dephasing, and nonlinear coupling.

The paper is organized as follows. In Sec. II, we introduce the theoretical model for nonlinearly coupled charger-mediated quantum batteries. In Sec. III, we investigate driven quantum batteries, focusing on the dynamics of energy storage and work extraction, as well as the optimization of charging time in the presence of dephasing. In Sec. IV, we investigate quantum-battery charging enabled by nonlinear coupling in the absence of external driving. Finally, we summarize our findings and discuss future perspectives in Sec. V.

%================================================
%=========================
%\begin{figure*}[t]
%	\centering
%	\includegraphics[width=0.85\textwidth]{1.pdf}
%	\caption{Phase-space evolution of a coherent state in a two-step moving potential. (a) Evolution in the co-moving frame. (b) Corresponding evolution in the laboratory frame. Dashed (filled) circles indicate the state at the beginning (end) of each free-evolution step.}\label{Fig1}
%\end{figure*}
%========================
%================================================

\section{Theoretical model}

\begin{figure}[t]
	\centering
	\includegraphics[width=0.48\textwidth]{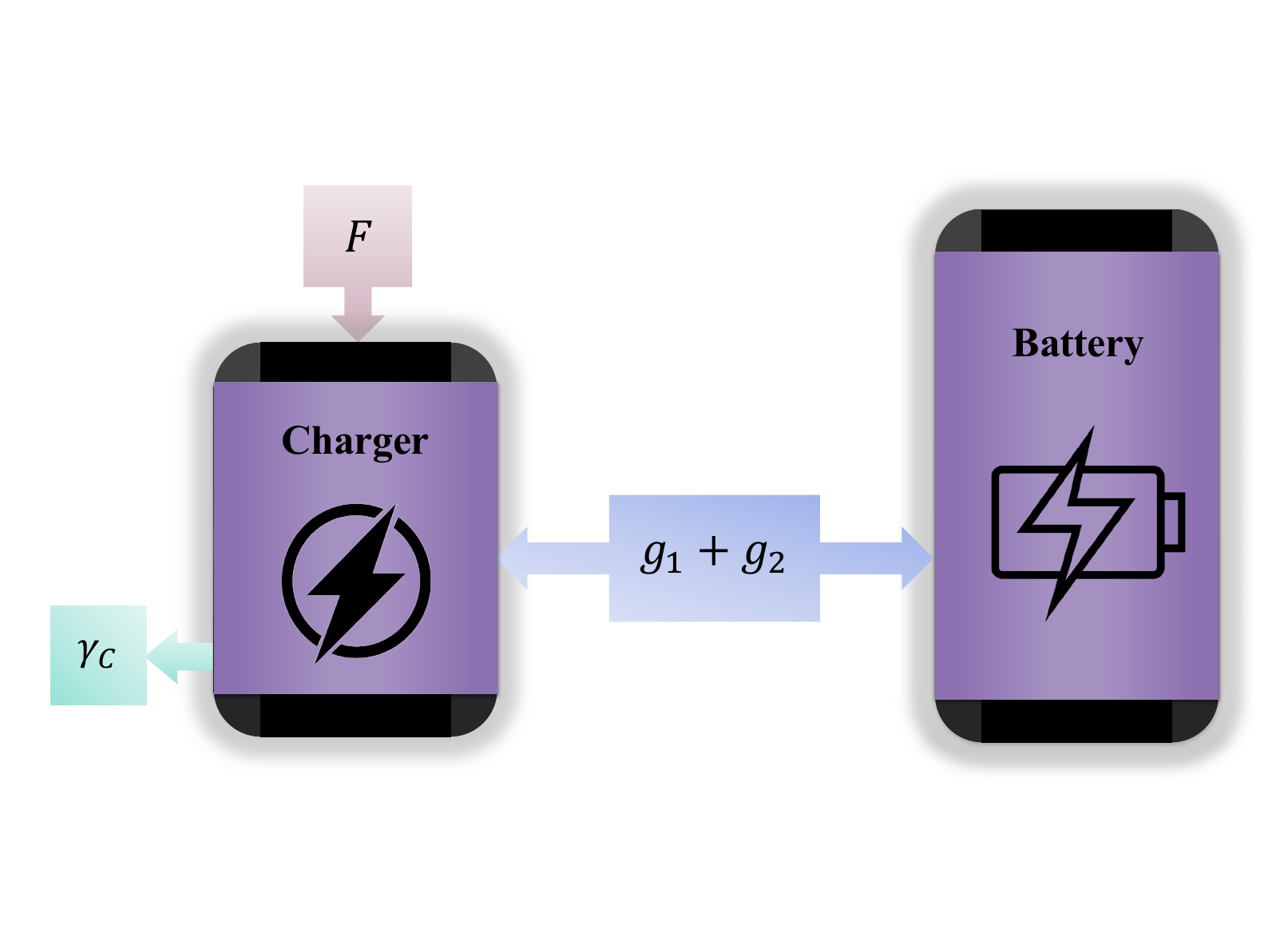}
	\caption{Schematic of the charger-battery setup with nonlinear coupling. The setup of a 
		quantum battery (B)is coupled to a charger (C), mediated by a combination of linear ($g_1$) and nonlinear ($g_2$) coupling terms. In addition, the charger is subject to an external coherent driving strength $F$ and a dephasing rate $\gamma_{\rm C}$.}\label{Fig1}
\end{figure}

We consider a charger-mediated quantum battery formed by a charger (C) system and a quantum battery (B) \cite{Campaioli2017}, as shown in Fig. \ref{Fig1}. The total Hamiltonian, including linear (L) and nonlinear (NL) interactions as well as coherent driving (D), can be written as
\bea
\hat H(t)= \hat H_{\rm C}+\hat H_{\rm B}+\hat H_{\rm L}+\hat H_{\rm NL}+\hat H_{\rm D}, 
\eea
where the intrinsic energy levels for the charger and battery systems are $\hat H_{\rm C(B)}=\omega_{\rm C(B)}\sigma_{\rm C(B)}^\dag\sigma_{\rm C(B)}$, respectively, $\hat H_{\rm L}= g_1(\sigma_{\rm C}^\dag\sigma_{\rm B}+ \sigma_{\rm B}^\dag\sigma_{\rm C})$, $\hat H_{\rm NL}= g_2(\sigma_{\rm C}^\dag\sigma_{\rm B}^\dag+ \sigma_{\rm C}\sigma_{\rm B})$, and the coherent driving term is $\hat H_{\rm D}=F(\sigma_{\rm C}e^{i\omega_{\rm D}t}+\sigma_{\rm C}^\dag e^{-i\omega_{\rm D}t})$. The raising and lowering operators of the two-level charger and battery are denoted by $\hat{\sigma}_{\rm C(B)}^\dagger$ and $\hat{\sigma}_{\rm C(B)}$, respectively. Throughout this work, we assume the resonance condition $\omega_{\rm C}=\omega_{\rm B}=\omega_{\rm D}$ among the charger, battery, and coherent driving field. We then transform the Hamiltonian into the interaction picture defined by $\hat{U}=\exp[-i(\hat{H}_{\rm C}+\hat{H}_{\rm B})t]$, such that $\hat{H}_{I}=\hat{U}^\dagger\hat{H}\hat{U}-i\hat{U}^\dagger\partial_t\hat{U}$. Under the resonance condition, the free evolution associated with the bare transition frequencies $\omega_{\rm C(B)}$ can be eliminated, and the explicit time dependence of the coherent driving terms is also removed. The remaining time-dependent contributions arise from the nonlinear interactions, with oscillatory factors proportional to $e^{\pm 2i\omega_{\rm B}t}$.

%================================================
%========================

Motivated by the dephasing-enabled quantum battery setup \cite{Shastri2025}, the charger system is subject to dephasing effect, and the density matrix of the joint system reads ($\hbar=1$)
\bea
\dot{\hat \rho}= -i[\hat H_I, \hat \rho(t)]+\gamma_{\rm C}\left(\hat L_{\rm C}\hat \rho(t)\hat L_{\rm C}-\frac{\{\hat L_{\rm C}^2,\hat \rho(t)\}}{2}\right), \label{rho}
\eea
where $\{.,.\}$ denotes the anticommutator, $\hat L_{\rm C}\equiv \hat \sigma_{\rm C}^\dag\hat \sigma_{\rm C}$, and $\gamma_{\rm C}$ quantifies the strength of the dephasing rates. We note that the two-level systems constituting the charger and quantum battery can be realized using two hyperfine ground states of neutral atoms or artificial atoms such as superconducting qubits \cite{Naghiloo2019, Chu2025}. In such platforms, the linear charger–battery coupling strength $g_1$ can become comparable to the level splitting $\omega_{\rm C(B)}$, providing access to a regime in which nonlinear couplings $g_2$ can play a significant role in the charging dynamics. These nonlinear couplings enable additional channels for energy exchange and transfer between the charger and the battery, with explicitly time-dependent contributions, which becomes relevant beyond the rotating-wave approximation. By contrast, the effect of $g_2$ becomes negligible in the weak linear-coupling regime, $\gamma_{\rm C},g_1\ll\omega_{\rm C(B)}\approx g_2$ \cite{Shastri2025}. In this limit, terms proportional to $e^{i2\omega_{\rm C(B)}t}$ associated with the nonlinear coupling process in Eq. (\ref{rho}) oscillate rapidly compared with the relevant system dynamics and therefore average out, rendering their contribution negligible \cite{Scully1997, Lehmberg1970, Jen2025_review}.

From Eq. (\ref{rho}), we obtain the self-consistent coupled equations of motion as 
\bea
\frac{d}{dt} \braket{\hat \sigma_{\rm B}^z}=&& -4g_1\operatorname{Im}[\braket{\hat \sigma_{\rm C}^\dag\hat\sigma_{\rm B}}]+4g_2\operatorname{Im}[e^{i2\omega_{\rm B}t}\braket{\hat \sigma^\dag_{\rm C}\hat\sigma^\dag_{\rm B}}], \nonumber \\\\
\frac{d}{dt} \braket{\hat \sigma_{\rm C}^z}=&& -4F\operatorname{Im}[\braket{\hat \sigma_{\rm C}}]+4g_1\operatorname{Im}[\braket{\hat \sigma_{\rm C}^\dag\hat\sigma_{\rm B}}]\nonumber\\&&+4g_2\operatorname{Im}[e^{i2\omega_{\rm B}t}\braket{\hat \sigma^\dag_{\rm C}\hat\sigma^\dag_{\rm B}}], \\
\frac{d}{dt} \braket{\hat \sigma_{\rm C}^\dag\hat\sigma_{\rm B}}=&&-iF\braket{\hat \sigma_{\rm C}^z\hat\sigma_{\rm B}}-\frac{\gamma_{\rm C}}{2} \braket{\hat \sigma_{\rm C}^\dag\hat\sigma_{\rm B}}\nonumber\\&&-i\frac{g_1}{2}(\braket{\hat \sigma_{\rm C}^z}-\braket{\hat \sigma_{\rm B}^z}), \\
\frac{d}{dt} \braket{\hat \sigma_{\rm C}^z\hat\sigma_{\rm B}}=&&-i2F(\braket{\hat \sigma_{\rm C}^\dag\hat\sigma_{\rm B}}-\braket{\hat \sigma_{\rm C}\hat\sigma_{\rm B}})+ig_1 \braket{\hat \sigma_{\rm C}}\nonumber\\&&-ig_2e^{i2\omega_{\rm B}t}\braket{\hat \sigma_{\rm C}^\dag}, \\
\frac{d}{dt} \braket{\hat \sigma_{\rm C}\hat\sigma_{\rm B}}=&&iF \braket{\hat \sigma_{\rm C}^z\hat\sigma_{\rm B}}-\frac{\gamma_{\rm C}}{2}\braket{\hat \sigma_{\rm C}\hat\sigma_{\rm B}}\nonumber\\&&+i\frac{g_2}{2}(\braket{\hat \sigma_{\rm C}^z}+\braket{\hat \sigma_{\rm B}^z}),\label{eq7}\\
\frac{d}{dt} \braket{\hat \sigma_{\rm C}}=&&iF\braket{\hat \sigma_{\rm C}^z}+ig_1 \braket{\hat \sigma_{\rm C}^z\hat\sigma_{\rm B}}-\frac{\gamma_{\rm C}}{2}\braket{\hat \sigma_{\rm C}}\nonumber\\&&+ ig_2\braket{\hat \sigma_{\rm C}^z\hat\sigma_{\rm B}^\dag}, \\
\frac{d}{dt} \braket{\hat \sigma_{\rm B}}=&&ig_1\braket{\hat \sigma_{\rm C}^z\hat\sigma_{\rm B}}+ig_2 \braket{\hat\sigma_{\rm C}^\dag\hat \sigma_{\rm B}^z}, \\
\frac{d}{dt} \braket{\hat\sigma_{\rm C}\hat \sigma_{\rm B}^z}=&&iF\braket{\hat \sigma_{\rm C}^z\hat \sigma_{\rm B}^z}+ig_1\braket{\hat \sigma_{\rm B}}-\frac{\gamma_{\rm C}}{2}\braket{\hat\sigma_{\rm C}\hat \sigma_{\rm B}^z}\nonumber\\&&-ig_2e^{i2\omega_{\rm B}t}\braket{\hat \sigma_{\rm B}^\dag} , \\
\frac{d}{dt} \braket{\hat \sigma_{\rm C}^z\hat \sigma_{\rm B}^z}=&&-4F\operatorname{Im}[\braket{\hat \sigma_{\rm C}^z\hat\sigma_{\rm B}}], 
\eea
where $\operatorname{Im}$ denotes the imaginary parts. The above equations reduce to those obtained in the linearly coupled regime \cite{Shastri2025} when $g_2=0$. The additional contributions arising from a finite $g_2$ coexist with the linear charging process governed by $g_1$, while introducing energy-nonconserving channels for energy exchange between the charger and the battery. The essential role of $g_2$ can be further identified from Eq. (\ref{eq7}), where the nonlinear charger–battery coupling provides a distinct mechanism for enhancing energy transfer, in addition to those enabled by coherent driving and dephasing. This nonlinear contribution therefore offers an additional resource for optimizing the charging performance of the quantum battery.

In the following sections, we focus on two common but important quantities that characterize the performance of quantum batteries: the stored energy and the ergotropy. The former quantifies the energy accumulated in the battery during the charging process, while its rate of change determines the corresponding charging power, i.e., the rate of energy transfer per unit time. The ergotropy, on the other hand, quantifies the maximum amount of work that can be extracted from a quantum state through unitary operations. It therefore characterizes the usable fraction of the energy stored in the quantum battery and provides an important metric of its capability for work extraction during the discharging process.

\section{Driven quantum battery with nonlinear couplings}

First in this section, we investigate the nonlinearly coupled quantum batteries under driven conditions, while the corresponding undriven scenario will be considered in the next section for comparison. As discussed at the end of the previous section, we employ two key quantities, the battery energy and ergotropy, to characterize the performance of the quantum battery. The battery energy is defined as $E_{\rm B}\equiv {\rm Tr}(\hat \rho_{\rm B}\hat H_{\rm B})$, where $\hat \rho_{\rm B}\equiv {\rm Tr}_{\rm C}(\hat \rho)$ denotes the reduced density matrix of the quantum battery obtained by tracing out the charger's degrees of freedom. The ergotropy, on the other hand, quantifies the maximum amount of work that can be extracted from a nonpassive state through unitary operations \cite{Lenard1978, Allahverdyan2004, Bhattacharjeea2021, Farina2019}, which can be defined as $\mathcal{E}_{\rm B}=E_{\rm B}-\min_{\hat U_{\rm B}}{\rm Tr}_{\rm B}(\hat U_{\rm B}\hat \rho\hat U_{\rm B}^\dag H_{\rm B})$. The $\hat U_{\rm B}$ denotes arbitrary unitary operators acting on the battery's Hilbert space, over which the minimization is performed. The ergotropy vanishes when $\hat \rho$ is a passive state, for which no further work can be extracted through cyclic unitary processes \cite{Allahverdyan2004}.

To further quantify how fast the charging process for energy transfer can be achieved, we adopt the definition of charging time used in the dephasing-enabled quantum batteries \cite{Shastri2025}, which is determined by
\bea
\left |\frac{E_{\rm B}(\tau)-E_{\rm B}(\infty)}{E_{\rm B}(0)-E_{\rm B}(\infty)}\right |=e^{-n}. \label{tau}
\eea
Here, $E_{\rm B}(\infty)$ denotes the steady-state battery energy, while a sufficiently large integer $n$ defines a characteristic charging time at which the battery energy approaches its steady-state value within a relative deviation bounded by $e^{-n}$.

As for the analytical forms of battery energy and ergotropy, they have been obtained as \cite{Farina2019}
\bea
E_{\rm B}=&&\frac{\omega_{\rm B}}{2}(\braket{\hat \sigma_{\rm B}^z}+1), \\
\mathcal{E}_{\rm B}=&&\frac{\omega_{\rm B}}{2}\left(\sqrt{\braket{\hat \sigma_{\rm B}^z}^2+4\braket{\hat \sigma_{\rm B}^\dag}\braket{\hat \sigma_{\rm B}}}+\braket{\hat \sigma_{\rm B}^z}\right).\label{ergo}
\eea
The energy of the quantum battery $E_{\rm B}$ can be calculated straightforwardly, which reduces to $E_{\rm B}=\omega_{\rm B}$ when the battery approaches the fully-charged state, that is $\braket{\hat \sigma_z}=1$, and to the vanishing $E_{\rm B}$ instead when the battery is in the empty state with $\braket{\hat \sigma_z}=-1$. The ergotropy of the battery can be obtained by assuming a generally Hermitian density matrix,
\bea
\hat\rho_g=&&
\begin{pmatrix}
	a & c^* \\
	c & b 
\end{pmatrix},\nonumber\\
=&&\frac{1}{2}[(a+b)\hat I_{\rm B}+(c+c^*)\hat \sigma_x+(c-c^*)\hat \sigma_y+(a-b)\hat \sigma_z],\nonumber\\
=&&\frac{1}{2}(I_{\rm B}+\vec r\cdot\vec\sigma),
\eea
where we redefine $\vec r\equiv(x,y,z)$ with corresponding combinations of arbitrary $a$, $b$, and $c$, and a vector of Pauli matrices $\vec\sigma\equiv(\hat \sigma_x,\hat\sigma_y,\hat\sigma_z)$. Then the eigenvalues of $\hat\rho_g$ can be calculated as $\lambda_{1,2}=\frac{1}{2}\pm|\vec r|$, which leads to the passive energy, $\min(\lambda_i E_i)=(1/2-|\vec r|)\omega_{\rm B}$ with $E_i\in \{0, \omega_{\rm B}\}$ and $i\in\{1,2\}$. The ergotropy of the battery becomes $\mathcal{E}_{\rm B}=E_{\rm B}-(1/2-|\vec r|)\omega_{\rm B}$, and leads to Eq. (\ref{ergo}). 

\subsection{Fast charging in nonlinear-coupled quantum battery}

Fast charging dynamics in the linear-coupling regime has been demonstrated in dephasing-enabled quantum batteries \cite{Shastri2025}. At small dephasing rates $\gamma_{\rm C}$, the battery exhibits transient overcharging associated with underdamped dynamics, whereas increasing $\gamma_{\rm C}$ eventually drives the system into an overdamped regime and slows down the charging process. The crossover between these two regimes therefore gives rise to an optimal dephasing rate for minimizing the charging time. In the long-time limit, the battery approaches the steady-state energy $E_{\rm B}(t\rightarrow\infty)=\omega_{\rm B}/2$ and ergotropy
$\mathcal{E}_{\rm B}(t\rightarrow\infty)=\omega_{\rm B}(F/g_1)/(1+4F^2/g_1^2)$ in the linear-coupling regime \cite{Shastri2025, Farina2019}. The latter further identifies an optimal coherent driving strength, $F/g_1=0.5$, at which the steady-state ergotropy reaches its maximum value, $\mathcal{E}_{\rm B}(\infty)=\omega_{\rm B}/4$. These results establish the optimal dephasing rate and driving strength as useful benchmarks for assessing the charging enhancement induced by nonlinear charger–battery coupling.

\begin{figure}[t]
	\centering
	\includegraphics[width=0.49\textwidth]{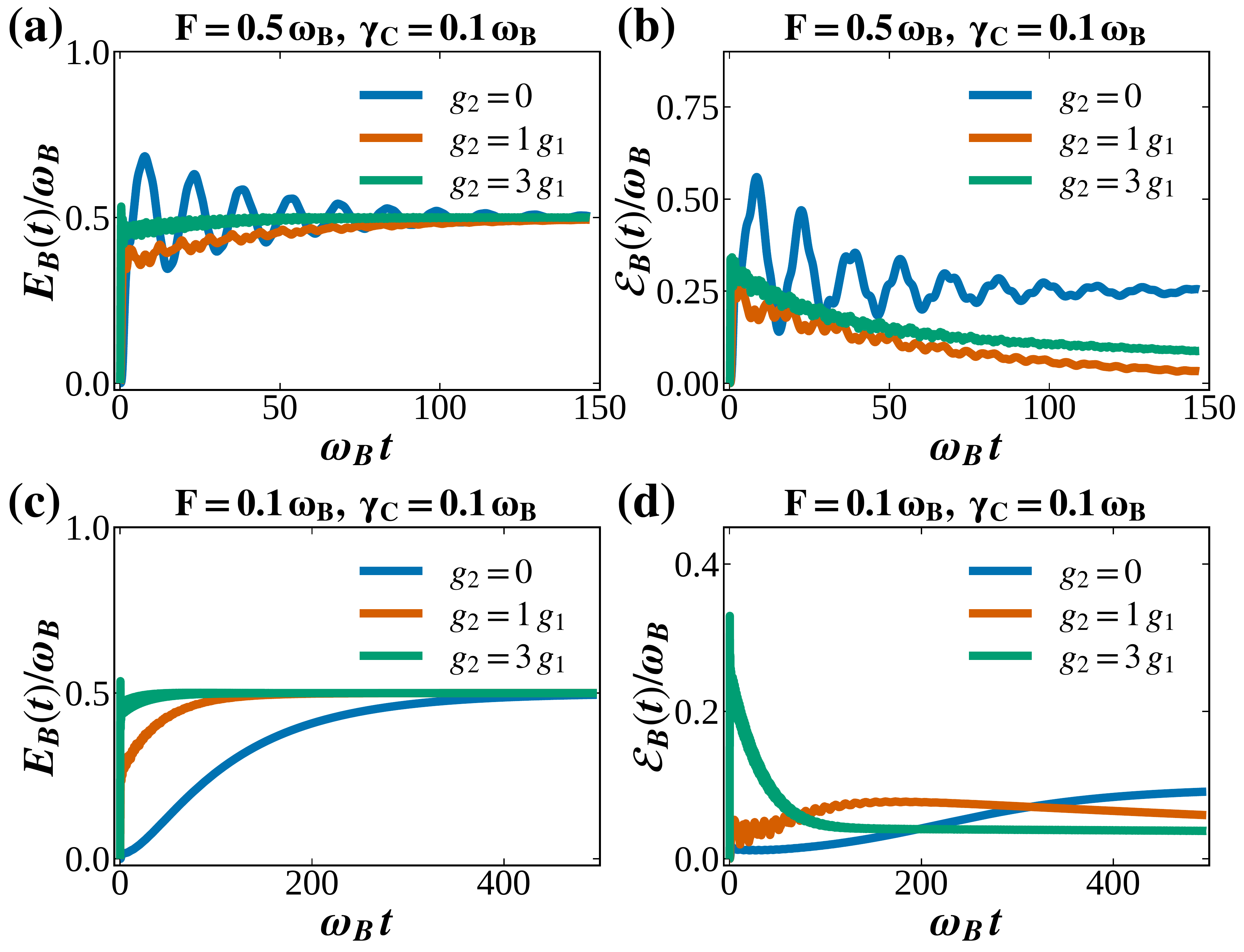}
	\caption{Time evolution of battery energy and ergotropy under nonlinear couplings. The dynamics of energy $E_{\rm B}$ and ergotropy $\mathcal{E}_{\rm B}$ respectively for (a,b) $F=0.5\omega_{\rm B}$ and (c, d) $F=0.1\omega_{\rm B}$, with the same dephasing rate $\gamma_C=0.1\omega_{\rm B}$ under linear-coupling regime $g_2=0$ and a regime with finite nonlinear couplings. Critically damped behaviors can be observed as $g_2$ increases, accelerating the charging dynamics.}\label{Fig2}
\end{figure}

In Fig. \ref{Fig2}, we investigate two driving regimes in the presence of finite nonlinear coupling $g_2$: the ergotropy-optimal driving regime of the linear-coupling system and a weaker-driving regime that requires less external driving resources. In the former case in Fig. \ref{Fig2}(a), with $F/g_1=0.5$, the introduction of finite $g_2$ drives the charging dynamics toward a nearly critically damped behavior, in contrast to the dynamics observed in the linear-coupling regime. Consequently, the battery approaches its steady state with strongly suppressed oscillations and on a shorter timescale. As $g_2$ increases, this nearly critically damped behavior is accompanied by a further enhancement of the charging rate, demonstrating the capability of nonlinear coupling to accelerate energy storage.

The behavior of the ergotropy $\mathcal{E}_{\rm B}$, however, reveals a distinct trade-off. As shown in Fig. \ref{Fig2}(b), for finite $g_2$, $\mathcal{E}_{\rm B}$ approaches its steady state on a timescale comparable to that in the linear-coupling regime ($g_2=0$), but reaches a lower steady-state value. This indicates that nonlinear coupling, while accelerating energy storage, can be detrimental to the steady-state extractable work. Nevertheless, $\mathcal{E}_{\rm B}$ can attain substantially larger transient values during the early stage of charging, approaching the steady-state ergotropy achieved in the linear-coupling regime. Moreover, its nearly critically damped evolution suppresses temporal oscillations, providing a relatively broad time window over which a large transient ergotropy can be extracted. The nonlinear-coupling regime therefore exhibits a trade-off between faster charging and reduced steady-state ergotropy, while retaining substantial extractable work at intermediate charging times.

%================================================
%========================

For comparison, in Figs. \ref{Fig2}(c) and \ref{Fig2}(d) we consider a weaker-driving regime at the same dephasing rate $\gamma_{\rm C}$. Increasing $g_2$ drives the initially overdamped charging dynamics toward a nearly critically damped regime and, similarly, substantially accelerates the approach of the battery energy to its steady state. Remarkably, a large transient ergotropy also emerges at early times and can even exceed the steady-state ergotropy of the corresponding linear-coupling system. These results demonstrate that nonlinear coupling can accelerate battery charging under both driving conditions, while the resulting energy-storage and work-extraction characteristics are determined by the interplay among coherent driving, nonlinear coupling, and dephasing.

We further note that the ergotropy is strongly suppressed at large dephasing rates in the presence of nonlinear coupling, even under finite coherent driving. This behavior can be attributed to the degradation of charger–battery correlations generated by the nonlinear interaction, which reduces the amount of extractable work stored in the battery. Nevertheless, the substantial reduction in charging time and the lower external driving requirement introduce an important operational trade-off between charging speed, driving cost, and extractable work. In particular, the shorter charging cycle enabled by nonlinear coupling may allow more frequent charging-and-discharging operations, potentially compensating for the reduced work extracted per cycle through an enhanced rate of work extraction. Motivated by this trade-off, in the next subsection we focus on the additional control provided by nonlinear coupling and explore its capability to enable fast charging over a broader range of system parameters.

%================================================
\begin{figure*}[t]
	\centering
	\includegraphics[width=0.95\textwidth]{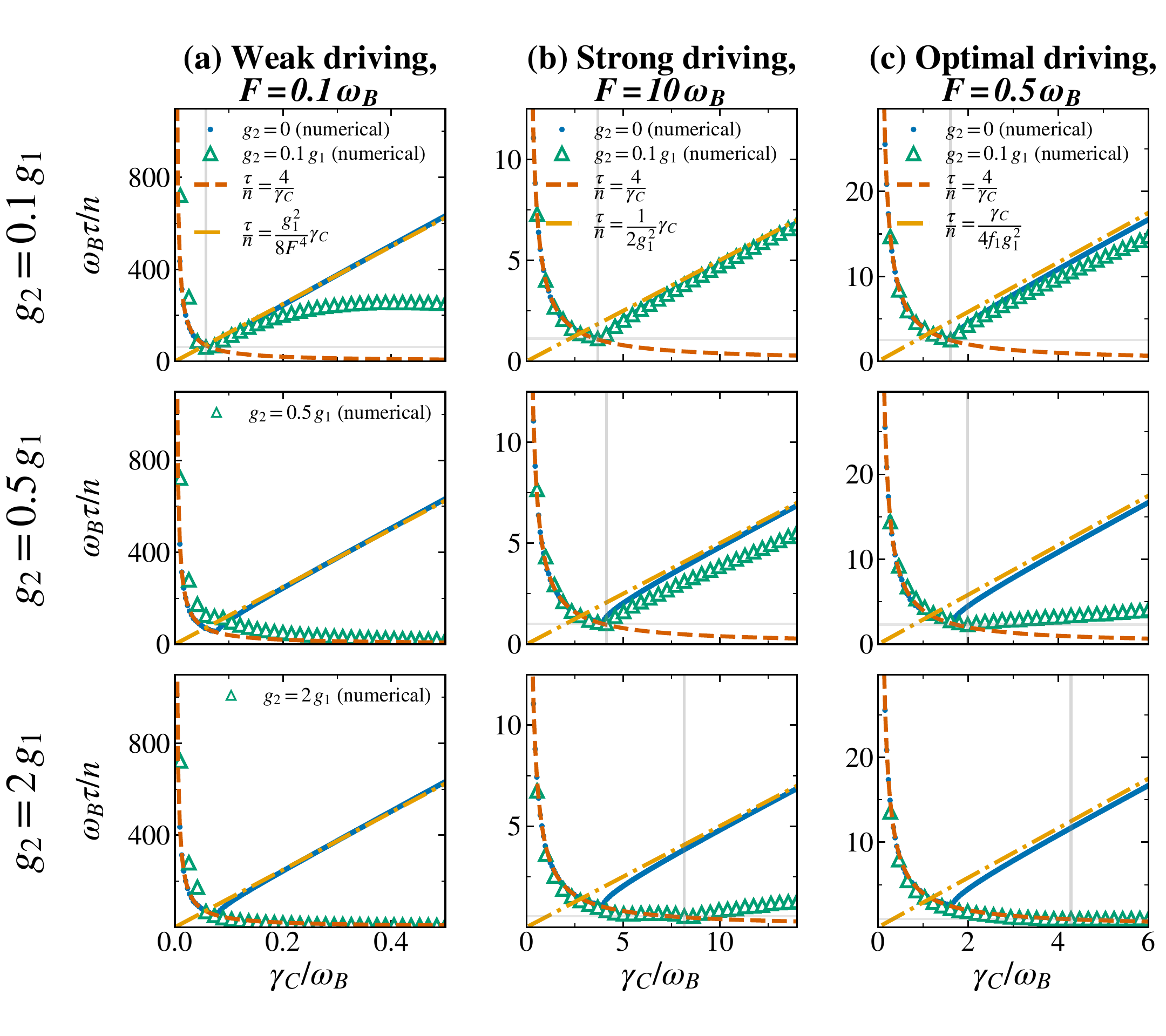}
	\caption{Plot of charging time ($\tau$) as a function of $\gamma_{\rm C}$ with different $g_2$. Rescaled charging time $\tau$ versus charger dissipation rate $\gamma_{\rm C}$ under (a) weak driving ($F/g_1=0.1$), (b) strong driving ($F/g_1=10$), and (c) optimal driving ($F/g_1=0.5$), with $g_1=\omega_{\rm B}$ and $n=18$ throughout. In each panel, blue dots show the $g_2=0$ baseline, while green triangles show the numerical results for finite $g_2$, taking the values $g_2=0.1\omega_{\rm B}$ (top row), $g_2=0.5\omega_{\rm B}$ (middle row), and $g_2=2\omega_{\rm B}$ (bottom row). The vertical and horizontal gray lines define the optimal dephasing rate ($\gamma_{\rm C}^*$) that minimizes the charging time for each numerical result with different $g_2$. As $g_2$ increases, the charging time deviates significantly from that in the linear regime ($g_2=0$), particularly at large $\gamma_{\rm C}$. This deviation becomes most pronounced for $g_2=2\omega_{\rm B}$, where the charging time is substantially reduced and eventually saturates as $\gamma_{\rm C}$ increases. The expression of scaling function $f_1$ can be found in the main text.}\label{Fig3}
\end{figure*}
%========================

\subsection{Improved charging time under nonlinear couplings}

To further characterize and quantify the charging time $\tau$ across different parameter regimes of $F$, $\gamma_{\rm C}$, and $g_2$, we adopt the definition in Eq. (\ref{tau}) with $n=18$ and present the corresponding results in Fig. \ref{Fig3}. Before discussing the numerical results, it is useful to recall several analytical estimates available in the linear-coupling regime \cite{Shastri2025}, which provide useful benchmarks for assessing the effects of nonlinear coupling. Throughout this work, we set $g_1=\omega_{\rm B}$, such that the linear-coupling strength and the battery energy scale provide a common reference for comparing the charging dynamics.

In the weak-dephasing limit, $\gamma_{\rm C}\ll g_1=\omega_{\rm B}$, the characteristic charging time is governed by the slow decay envelope $e^{-\gamma_{\rm C}t/4}$ appearing in $\braket{\hat{\sigma}_{\rm B}^z(t)}$ and, consequently, in $E_{\rm B}(t)$. The corresponding decay timescale is independent of $F$ and $g_1$, yielding from Eq. (\ref{tau}) a useful time scale of $\tau\sim 4n/\gamma_{\rm C}$. In contrast, in the strong-dephasing regime, $\gamma_{\rm C}\gg F,g_1$, the charging time acquires an explicit dependence on the driving ratio $F/g_1$ \cite{Shastri2025}. By identifying the slowest decay timescale governing the approach to the steady state, one obtains $\tau\sim n g_1^2\gamma_{\rm C}/(8F^4)$ for $F/g_1\ll 1$ in the weak driving limit, while $\tau\sim n \gamma_{\rm C}/(2 g_1^2)$ for $F/g_1\gg 1$ in the strong driving limit. These asymptotic expressions provide useful reference timescales against which the charging enhancement induced by finite nonlinear coupling $g_2$ can be quantified.  

In Fig. \ref{Fig3}, we investigate three representative driving regimes: weak driving in (a), strong driving in (b), and optimal driving in (c), where the latter corresponds to the driving strength that maximizes the steady-state ergotropy $\mathcal{E}_{\rm B}$ in the linear-coupling regime and serves as a reference for comparison. In the weak-driving regime, a small nonlinear coupling $g_2$ does not further reduce the charging time $\tau$ at the optimal dephasing rate $\gamma_{\rm C}^*$ of the linear-coupling system. Nevertheless, at larger $\gamma_{\rm C}$, a finite $g_2$ can substantially reduce $\tau$ compared with the linear-coupling case, $g_2=0$, revealing the potential benefit of combining dephasing with nonlinear charger–battery coupling.

As $g_2$ is further increased in Fig. \ref{Fig3}(a), the region of efficient charging shifts toward relatively large dephasing rates, where a pronounced reduction in $\tau$ emerges. Interestingly, in the weak-driving and strong-nonlinear-coupling regime, the charging time approximately follows the scaling $\tau\sim4n/\gamma_{\rm C}$. For example, the optimal charging time in the linear-coupling regime, $\omega_{\rm B}\tau/n\approx50$ at $\gamma_{\rm C}^*/\omega_{\rm B}\approx 0.08$, can be reduced to $\omega_{\rm B}\tau/n<10$ already at $\gamma_{\rm C}/\omega_{\rm B}=0.5$ for finite $g_2$, corresponding to a reduction in charging time by more than a factor of five. This substantial enhancement indicates that sufficiently strong nonlinear coupling can overcome the charging slowdown otherwise induced by dephasing and extend the regime of dephasing-enabled fast charging. In particular, the most pronounced enhancement occurs when $g_2\gtrsim g_1\gg F$, identifying strong nonlinear charger-battery coupling as the dominant resource for fast charging in the weak-driving regime.

In contrast, in the strong-driving regime shown in Fig. \ref{Fig3}(b), the reduction in charging time is less pronounced. Even for nonlinear coupling strengths $g_2$ comparable to those considered in Fig. \ref{Fig3}(a), the decrease in $\tau$ follows a trend similar to that observed in the weak-driving regime only up to $\gamma_{\rm C}/\omega_{\rm B}\approx10$ when $g_2=2g_1$. Beyond this regime, dephasing becomes dominant and $\tau$ increases again. Nevertheless, the interplay between nonlinear coupling and dephasing gives rise to a new optimum at $\gamma_{\rm C}^*/\omega_{\rm B}\approx8$, where the charging time is reduced by approximately a factor of two compared with the optimal value in the linear-coupling regime.

Finally, for the ergotropy-optimal driving scheme shown in Fig. \ref{Fig3}(c), the charging time $\tau$ is again progressively reduced as $g_2$ increases. In particular, for sufficiently large $g_2$, $\tau$ approaches the weak-dephasing scaling $\tau\sim4n/\gamma_{\rm C}$ over an extended range of the considered dephasing rates, before reaching a new optimum at $\gamma_{\rm C}^*/\omega_{\rm B}\approx4.2$ when $g_2=2g_1$. At this optimum, the minimum charging time is reduced by approximately a factor of three relative to that in the linear-coupling regime. Importantly, both this driving scheme and weak-driving scheme in Fig. \ref{Fig3}(a) can simultaneously allow transient ergotropy of the quantum battery. These results demonstrate that sufficiently strong nonlinear charger–battery coupling can significantly shorten the charging time and provide a promising route toward quantum batteries with both rapid charging and high work-extraction capability. We note that as a comparison in the linear-coupling regime in Fig. \ref{Fig3}(c), we employ the analytical estimate $\tau\sim n\gamma_{\rm C}/(4 f_1 g_1^2)$ for $F/g_1\lesssim 1$ where $f_1\equiv 1+2F^2/g_1^2-\sqrt{1+4F^2/g_1^2}$ \cite{Shastri2025} to correctly describe the asymptotic behavior of $\tau$ at relatively large $\gamma_{\rm C}$. 

\section{Undriven nonlinear coupled quantum battery}

%================================================
\begin{figure}[t]
	\centering
	\includegraphics[width=0.48\textwidth]{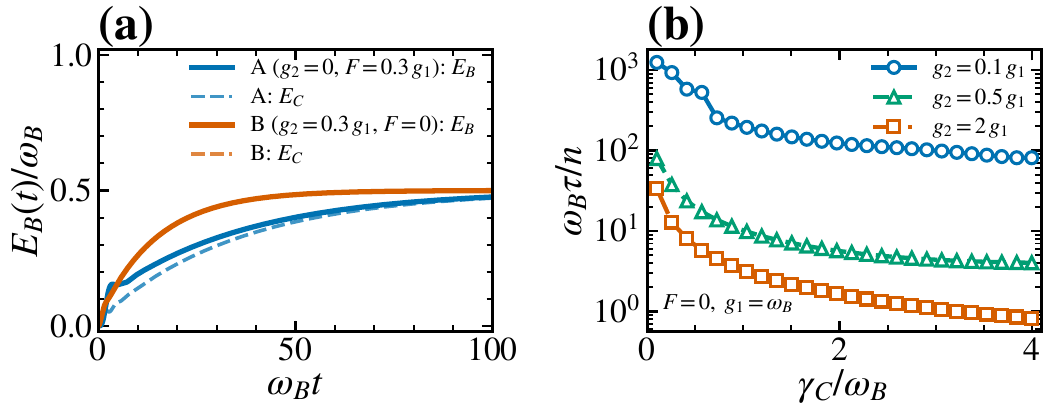}
	\caption{Dynamics of energy and charging time for a quantum battery with nonlinear coupling without external driving. (a) Time evolution of the normalized battery ($E_{\rm B}$) and charger ($E_{\rm C}$) energies under the resonance condition $\omega_{\rm B}=\omega_{\rm C}=\omega_D$, with $\gamma_{\rm C}=2\omega_{\rm B}$ and $g_1=\omega_{\rm B}$. We compare protocol A (blue), corresponding to classical driving with $F/g_1=0.5$, and protocol B (orange), corresponding to nonlinear coupling without external driving, with $g_2/g_1=0.5$. While both protocols approach the same steady-state energy, $E_{\rm B}=E_{\rm C}=0.5\omega_{\rm B}$, protocol B exhibits significantly faster initial charging dynamics, highlighting the charging enhancement enabled by nonlinear coupling alone. Note that $E_{\rm C}(t)=E_{\rm B}(t)$ in this case. (b) Charging time $\tau$ as a function of the dephasing rate $\gamma_{\rm C}$ in the undriven nonlinear-coupling regime ($F=0$ and $g_1=\omega_{\rm B}$) for three nonlinear coupling strengths, $g_2/g_1=0.1$, $0.5$, and $2$. Increasing $g_2$ reduces the charging time by almost two orders of magnitude, demonstrating that nonlinear coupling without external driving can activate a self-charging mechanism and serve as an effective resource for accelerating battery charging.}\label{Fig4}
\end{figure}
%========================

Next, we turn off the external driving field, $F=0$, and focus on quantum-battery charging enabled by nonlinear coupling only. Although coherent driving and nonlinear coupling $g_2$ play distinct roles in the charging dynamics, it is natural to ask whether a finite $g_2$ alone can outperform the conventional linear-coupling scheme assisted by coherent driving. To address this question, in Fig. \ref{Fig4}(a) we compare the two schemes at equal strengths for coupling and driving, $g_2=F$, respectively. For $g_2/\omega_{\rm B}=F/\omega_{\rm B}\lesssim0.4$, the undriven nonlinear-coupling scheme exhibits a clear reduction in the charging time relative to its driven linear-coupling counterpart. As $g_2=F$ is further increased, however, this advantage gradually disappears. This crossover is governed by the dephasing rate $\gamma_{\rm C}$: for a given $\gamma_{\rm C}$, nonlinear coupling provides a charging advantage only up to a characteristic coupling strength, beyond which coherent driving becomes more efficient. As expected, increasing $\gamma_{\rm C}$ extends the parameter range over which the undriven nonlinear-coupling scheme remains advantageous. This behavior is consistent with Fig. \ref{Fig3}, where sufficiently strong nonlinear coupling can overcome the charging slowdown induced by dephasing and outperform the corresponding linear-coupling regime. We note that the charger energy evolves synchronously with that of the battery, $E_{\rm C}(t)=E_{\rm B}(t)$, demonstrating synchronous energy transfer enabled by nonlinear coupling. This behavior contrasts with the linear-coupling regime, where a slight mismatch between the charger and battery energies indicates a finite time delay in the energy-transfer process before both systems reach their steady states.

To further demonstrate the advantage of nonlinear coupling without external driving, in Fig. \ref{Fig4}(b) we investigate the charging time as a function of the dephasing rate $\gamma_{\rm C}$ for different values of $g_2$. As $g_2$ increases, the charging time is substantially reduced and can even reach $\omega_{\rm B}\tau/n<1$, surpassing the optimal charging timescale obtained for the driven linear-coupling scheme in Fig. \ref{Fig3}(c). This result demonstrates that external coherent driving is not essential for achieving rapid charging, that is, sufficiently strong nonlinear charger–battery coupling, assisted by an appropriate dephasing rate, can itself provide an efficient channel for energy transfer. Figure \ref{Fig4}(b) therefore highlights the potential of undriven nonlinear-coupling quantum batteries to achieve charging times beyond those accessible in the conventional driven linear-coupling regime.

\section{Conclusion and Discussion}

In this paper, we demonstrate that nonlinear charger–battery coupling provides an effective resource for further accelerating the charging of quantum batteries. For charger-mediated quantum batteries subject to dephasing, we uncover a nearly critically damped charging regime that gives rise to a distinct optimum in the charging time, highlighting the interplay between nonlinear coupling and dephasing as a mechanism for efficient energy transfer. For sufficiently strong nonlinear coupling, rapid charging can be achieved while simultaneously retaining finite steady-state ergotropy, thereby combining fast energy storage with work-extraction capability. Importantly, even in the absence of external driving, nonlinear coupling can significantly accelerate the charging dynamics, offering an alternative route toward efficient quantum energy storage. Our results establish nonlinear charger–battery coupling as a valuable resource for enhancing quantum-battery performance and provide new opportunities for designing quantum energy-storage devices with both rapid charging and high extractable work.

Our proposal can be implemented in state-of-the-art quantum platforms, with superconducting qubits \cite{Blais2021} and waveguide QED systems \cite{Sheremet2023, Lodahl2017} providing particularly promising settings. Superconducting circuits offer a highly controllable platform for realizing a broad class of open quantum-optical models, including Dicke-type quantum batteries \cite{Campaioli2024, Dicke1954}, as well as for exploring non-Hermitian quantum dynamics \cite{Naghiloo2019, Chen2021, Li2023, Feyisa2025_1, Feyisa2025_2}. Waveguide QED systems, on the other hand, provide access to strong light–matter interactions that can facilitate rapid and efficient energy transport \cite{Chen2025, Yu2025}. These platforms also enable precise control over quantum states and engineered interactions, including directional couplings \cite{Mitsch2014}, providing a versatile foundation for developing nanoscale quantum devices that integrate energy storage, transport, and quantum-information processing.

As an outlook, several extensions may further broaden the scope of quantum-battery architectures. Monitored quantum batteries \cite{Cenedese2026, Ferraro2026, Gherardini2020}, collectively coupled quantum batteries \cite{Shukla2026, Campaioli2017}, and quantum-battery networks \cite{Ferraro2026} provide natural settings for investigating enhanced work extraction, charging power assisted by multipartite quantum correlations, and energy-efficient quantum-information processing. In particular, extending the nonlinear interactions considered here to many-body battery architectures represents a promising direction. For example, star-topology configurations \cite{Joshi2022} may exploit such nonlinear charging processes to achieve performance beyond that attainable with correlated charger–battery clusters. Other important directions include exploring the competition among dynamical pathways generated by many-body interactions of different ranges \cite{Shukla2026}, quantifying the energetic costs associated with quantum measurements and feedback \cite{Ferraro2026}, and incorporating time-controlled dark-state energy storage in $V$-type three-level quantum batteries \cite{Chu2026}. Together, these directions may deepen our understanding of the fundamental mechanisms governing quantum-battery operation and advance the development of scalable and efficient quantum energy-storage devices.

\section*{Acknowledgments}

We acknowledge support from the National Science and Technology Council (NSTC), Taiwan, under the Grants No. NSTC-115-2112-M-001035-MY3 and No. NSTC-115-2119-M-001-009, and from Academia Sinica under Grant AS-CDA-113-M04. We are also grateful for support from TG 1.2 of NCTS at NTU.

\section*{Data Availability}

The data that support the findings of this article are not publicly available. The data are available from the authors upon reasonable request.
 
%%%%%%%%%%%%%%%%%%%%%%%%%%%%%%%%%%%

%%%%%%%%%%%%%%%%%%%%%%%%%%%%%%%%%%%%%%%%%%%%%%%%%%%%%%%%
\end{document}